\documentclass[12pt]{iopart}
\usepackage{graphics}
\usepackage{bm}
\usepackage{amssymb}
\usepackage{amsfonts}

\begin{document}

\title{Exact dynamics and squeezing in two harmonic modes  coupled through
angular momentum}

\author{N.\ Canosa$^{1}$, Swapan Mandal$^{1,2}$ and R.\ Rossignoli$^{1}$}

\address{
$^{1}$ Departamento de F\'{\i}sica -IFLP, F.C.E., Universidad Nacional de La Plata,
C. C. 67, 1900 La Plata, Argentina\\
$^{2}$Department of Physics, Visva-Bharati, Santiniketan-731235,
India }

\begin{abstract}
We investigate the exact dynamics of a system of two independent harmonic
oscillators coupled through their angular momentum. The exact analytic solution
of the equations of motion for the field operators is derived,  and the
conditions for dynamical stability are  obtained. As application, we examine
the emergence of squeezing and mode entanglement for an arbitrary separable coherent
initial state. It is shown that close to instability, the system develops
considerable entanglement, which is accompanied with simultaneous squeezing in
the coordinate of one oscillator and the momentum of the other oscillator. In
contrast, for weak coupling away from instability, the generated entanglement
is small, with weak  alternating squeezing in the coordinate and momentum of
each oscillator. Approximate  expressions describing these regimes are also
provided.
\end{abstract}

\maketitle

\section{Introduction}
Models based on coupled  harmonic oscillators have long attracted attention in
several different fields due to their wide range of applications
\cite{Louisell,Walls,Mollow,EKN.68,Iafrate,Holzwarth,Belkin,Cochrane,Fan,Ng}.
In particular, the case of two harmonic modes coupled through their angular
momentum, which describes the motion of  a charged particle within a general
harmonic trap in a uniform magnetic field  or, equivalently, the motion in a
rotating anisotropic harmonic potential \cite{Va.56,FK.70,RS.80}, has been
employed in distinct scenarios, such as rotating nuclei \cite{RS.80}, quantum
dots in a magnetic field \cite{MC.94} and fast rotating Bose-Einstein
condensates \cite{LNF.01,OO.04,AF.07,ABL.09} within the lowest Landau level
approximation \cite{ABD.05,BDZ.08,FF.09}. Since its Hamiltonian is  quadratic
in the field operators, the model is also suitable for simulation with optical
techniques \cite{PE.94}.

In a previous work \cite{RK.09} its dynamics in coordinate representation was
analyzed in detail, showing that it exhibits a complex dynamical phase diagram,
with stable as well as distinct types of unstable (i.e., unbounded) dynamics.
We have also examined the generated entanglement between the modes, both in
ground and thermal states \cite{RR.11} (vacuum and thermal entanglement) as
well as that obtained after starting from a separable vacuum state
\cite{RCR.14}. It was shown, in particular, that this system is able to mimic
typical entanglement growth regimes arising after a quantum quench in complex
many body scenarios \cite{SLRD.13}. Entanglement is of course essential for
quantum information applications \cite{NC.00}, and a large entanglement growth
with time after starting from a separable state in a many-body system, is
indicative of a system dynamics which cannot be efficiently simulated by classical
means.

On the other hand, quantum squeezing constitutes another topic of great current
interest \cite{DS.01,JE.08,GT.09,Ma.11}. Its relation with entanglement has
been investigated in different systems
\cite{DS.01,JE.08,GT.09,Ma.11,CB.05,TO.09,CH.10}, with entanglement normally
inducing squeezing in certain observables. In particular, in \cite{CB.05} the
exact dynamics of entanglement and squeezing in a two-mode Bose-Einstein
condensate interacting through a Josephson-like coupling was determined.
Squeezing is important for quantum metrology, i.e., for improving the accuracy
in quantum measurements \cite{Ma.11,CG.12}, and it has been shown that in some
cases spin squeezing can be employed to detect entanglement
\cite{DS.01,GT.09,Ma.11,TO.09}. Nonetheless, in some regimes (like the linear
case of the model considered in \cite{CB.05}) squeezing may also arise without
substantial entanglement.

In this  work  we first derive the exact analytic expressions for the temporal
evolution of the Heisenberg field operators of two harmonic modes coupled
through their angular momentum.   The obtained result is valid for all values of
the system parameters, i.e., in stable as well as unstable dynamical regimes,
and allows to determine the exact evolution of an arbitrary observable of the
system. We then apply this result to determine and examine the dynamics of
squeezing, which has so far not been investigated in this model, and its
relation with the generated entanglement, when starting from a separable
coherent initial state. We will show that different regimes can arise depending
on the value of the rotational frequency. Close to the  instability point,
appreciable entanglement is generated, accompanied with simultaneous squeezing
in one of the variables of each mode, while for small couplings, the generated
entanglement is weak, with small squeezing appearing in both variables of the
mode at alternating times. Approximate simple expressions describing these two
distinct regimes are also provided.

\section{Formalism}
\subsection{The model}
The Hamiltonian of the system under study can be written as
\begin{equation}
H=\frac{1}{2m}P_{1}^{2}+\frac{m\omega_{1}^{2}}{2}Q_{1}^{2}+\frac{1}{2m}P_{2}^{2}+
\frac{m\omega_{2}^{2}}{2}Q_{2}^{2}-\omega\left(Q_{1}P_{2}-P_{1}Q_{2}\right)\,,\label{1}
\end{equation}
where the subscripts $1$ and $2$ refer obviously to the first and second
oscillator. The oscillator frequencies $\omega_1$, $\omega_2$ and the rotation
frequency $\omega$ will be assumed real and satisfying, without loss of
generality,
\begin{equation}\omega_1\geq \omega_2>0\,,\;\;\;\omega\geq 0\,.\end{equation}
In the case of a  particle of charge $e$ in a magnetic field $\bm{H}$ along the
$z$ axis within a harmonic trap with spring constants $K_1$, $K_2$ in the $x,y$
plane,  we have $\omega=e|\bm{H}|/(2mc)$, with $\omega_j^2=K_j/m+\omega^2$
($j=1,2$) \cite{RK.09} (motion along $z$ is obviously  decoupled from that in
$x,y$ plane, which is that described by (\ref{1})). Eq.\ (\ref{1}) also
represents the intrinsic Hamiltonian describing the motion (in the $x,y$ plane)
of a particle in a harmonic trap with constants $m\omega_j^2$ rotating with
frequency $\omega$ around the $z$ axis \cite{RK.09}.

By expressing the position ($Q_j$)  and momentum ($P_j$) operators in terms of
the dimensionless annihilation ($a_{j}$) and creation $(a_{j}^{\dagger})$ boson
operators, $Q_{j}  = \sqrt{\frac{\hbar}{2m\omega_{j}}}(a_{j}+a_{j}^{\dagger})$,
$P_{j}  = -i\sqrt{\frac{\hbar m\omega_{j}}{2}}(a_{j}-a_{j}^{\dagger})$,
$j=1,2$, we can rewrite  (\ref{1}) as
\begin{eqnarray}
H&=&\hbar\omega_{1}\left(a_{1}^{\dagger}a_{1}+\frac{1}{2}\right)+\hbar\omega_{2}\left(a_{2}^{\dagger}a_{2}
+\frac{1}{2}\right)-i\hbar\lambda_{+}\left(a_{2}^{\dagger}a_{1}-a_{1}^{\dagger}a_{2}\right)\label{4a}\\
&&-i\hbar\lambda_{-}\left(a_{1}a_{2}-a_{1}^{\dagger}a_{2}^{\dagger}\right)\,,\label{4}
\end{eqnarray}
where
\begin{equation}
\lambda_{\pm}  =
\omega\left(\frac{\omega_1\pm\omega_2}{2\sqrt{\omega_1\omega_2}}\right)\,.
\label{5}
\end{equation}
In the isotropic case $\omega_1=\omega_2=\omega_0$,  $\lambda_-=0$ and both $H$
and the angular momentum, which becomes just $-i\hbar \lambda_+(a^\dagger_2
a_1-a_1^\dagger a_2)$, commute with the total boson number
$N=a^\dagger_1a_1+a^\dagger_2 a_2$ and between themselves.

However, in the anisotropic case $\omega_1\neq \omega_2$, the angular momentum
term does not commute with $H$ and no longer conserves the total boson number.
This entails, in particular, that in contrast with the isotropic case,  the
system vacuum will become entangled as $\omega$ increases, and that an
initially separable state $|0_1\rangle|0_2\rangle$ (product of the vacuum of
each oscillator) will become entangled as time increases if $\omega\neq 0$. In
addition, the system may become unstable if $\omega$ increases sufficiently, as
discussed below \cite{RK.09}. Note that the exact dynamics of the special case
$\omega_1=\omega_2$ with a $Q_1 Q_2$ coupling, which was studied in detail  in
\cite{EKN.68}, corresponds formally to the case $\lambda_+=\lambda_-$, after
replacing $a_1\rightarrow ia_1$. Also, the two-mode Bose Einstein condensate
model of \cite{CB.05} (which contains as well non-linear terms) would formally
correspond in the linear case to $\lambda_-=0$ and a time-dependent
$\lambda_+$.

\subsection{Exact solution}
Let us now derive the explicit solution of the Heisenberg
equations of motion for the field operators,
\begin{equation}
i\hbar\dot{a_{j}}=[a_{j},H]\,,\,\,\,\, j=1,2\,.\label{6}
\end{equation}
Eq.\ (\ref{6}) leads to the linear system
\begin{equation}
\begin{array}{lcl}
\dot{a_{1}} & = & -i\omega_{1}a_{1}+\lambda_{+}a_{2}+\lambda_{-}a_{2}^{\dagger}\\
\dot{a_{2}} & = & -i\omega_{2}a_{2}-\lambda_{+}a_{1}+\lambda_{-}a_{1}^{\dagger}
\end{array}\,,\label{7}
\end{equation}
which can be written in matrix form as
\begin{equation}
i\left(\begin{array}{c}
\dot{\bm{a}}\\\dot{\bm{a}^\dagger}\end{array}\right) = {\cal H}
\left(\begin{array}{c}
\bm{a}\\\bm{a}^\dagger\end{array}\right)\,,\;\;\;\label{8M}
\end{equation}
where $\bm{a}=\left(\begin{array}{c}a_1\\a_2\end{array}\right)$,
$\bm{a}^\dagger=\left(\begin{array}{c}a^\dagger_1\\a^\dagger_2\end{array}\right)$
and  ${\cal H}$ is the  $4\times 4$ non-hermitian matrix
\begin{equation}
{\cal H}=\left(\begin{array}{cccc}\omega_1&i\lambda_+&0&i\lambda_-\\
-i\lambda_+&\omega_2&i\lambda_-&0\\
0&i\lambda_-&-\omega_1&i\lambda_+\\i\lambda_-&0&-i\lambda_+&-\omega_2\end{array}\right)\,.
\label{M0}
\end{equation}
The exact solution of Eq.\  (\ref{8M}) can be expressed as
\begin{equation}
\left(\begin{array}{c}
\bm{a}(t)\\\bm{a}^\dagger(t)\end{array}\right)  = {\cal U}(t)
\left(\begin{array}{c}
\bm{a}(0)\\\bm{a}^\dagger(0)\end{array}\right)\,,\;\;
\label{M}
\end{equation}
where
\begin{equation}
{\cal U}(t)=\exp[-i{\cal H} t]=\left(\begin{array}{cc}U(t)&V(t)\\
V^*(t)&U^*(t)\end{array}\right)\,,\label{exp}
\end{equation}
is a $4\times 4$ matrix satisfying ($\mathbb{I}$ denotes the $2\times 2$
identity matrix)
\begin{equation}{\cal U}(t){\cal M}{\cal U}^\dagger(t)={\cal M},\;\;{\cal M}
=\left(\begin{array}{cc}\mathbb{I}&0\\0&-\mathbb{I}\end{array}\right)
\label{Un}\,,\end{equation}
 since ${\cal M}{\cal H}^\dagger{\cal M}={\cal H}$.
Eq.\ (\ref{Un}) ensures the preservation of the equal time commutation
relations $\forall$ $t$:
\begin{equation}[a_i(t),a_j^\dagger(t)]=[UU^\dagger-VV^\dagger]_{ij}(t)=\delta_{ij}\,,
\;\;[a_i(t),a_j(t)]=[UV^{\rm t}-VU^t]_{ij}(t)=0\,,\label{com}\end{equation}
implying that Eq.\ (\ref{M}) represents a proper time-dependent Bogoliubov
transformation for the field operators.

Setting in what follows $a_j(0)\equiv a_j$, $a^\dagger_j(0)\equiv a^\dagger_j$,
Eq.\ (\ref{M}) leads explicitly  to
\begin{equation}
\begin{array}{lcl}
a_{j}(t) & = &
U_{j1}(t)\,a_{1}+U_{j2}(t)\,a_{2}+V_{j1}(t)\,a_{1}^{\dagger}+V_{j2}(t)\,a_{2}^{\dagger}\,,
\end{array}\label{8}
\end{equation}
and the corresponding adjoint equations for $a^\dagger_j(t)$, where the
elements $U_{jk}(t)$, $V_{jk}(t)$ can be obtained from Eq.\ (\ref{exp}) through
the diagonalization of ${\cal H}$:
\begin{eqnarray}
U_{jj}(t)&=&\frac{1}{2}\{(1+\gamma_j)\cos\omega_+ t
+(1-\gamma_j)\cos\omega_-t\nonumber\\&&-i\omega_j[(1+\delta_j)\frac{\sin\omega_+t}{\omega_+}+
(1-\delta_j)\frac{\sin\omega_- t}{\omega_-}]\}\,,\label{sol0}\\
V_{jj}(t)&=&i(-1)^{j+1}\frac{\omega_1\omega_2\lambda_+\lambda_-}{\omega_j\,\Delta}
(\frac{\sin\omega_+ t}{\omega_+}-\frac{\sin\omega_- t}{\omega_-})\,,\\
\left(\begin{array}{c}U_{12}(t)\\V_{12}(t)\end{array}\right)&=&
\frac{\lambda_{\pm}}{2}[(1+\frac{(\omega_1\pm    \omega_2)^2}{2\Delta})
\frac{\sin\omega_+t}{\omega_+}+(1-\frac{(\omega_1\pm\omega_2)^2}{2\Delta})
\frac{\sin\omega_-t}{\omega_-}\nonumber\\&&+i\frac{\omega_1\pm\omega_2}{\Delta}
(\cos\omega_+ t-\cos\omega_- t)]\,,\\
U_{21}(t)&=&-U_{12}(t)\,,\;\;V_{21}(t)=V_{12}^*(t)\,,\label{solf}
\end{eqnarray}
 with
 \begin{equation}\gamma_j=(-1)^{j+1}\frac{\omega_1^2-\omega_2^2}{2\Delta},\;\;
\delta_j=\gamma_j+\frac{\omega^2(2\omega_j^2+\omega_1^2+\omega_2^2)}{2\Delta\omega_j^2}\,.
\label{delj}\end{equation}
Here $\omega_{\pm}$ are the system {\it eigenfrequencies}, i.e., the eigenvalues of
the matrix ${\cal H}$ (which are  $\pm\omega_+$, $\pm\omega_-$), given by
\begin{eqnarray}\omega_{\pm}&=&\sqrt{\lambda_+^2-\lambda_-^2+{\textstyle\frac{\omega_1^2+\omega_2^2}{2}}
\pm\Delta}=\sqrt{\omega^2+{\textstyle\frac{\omega_1^2+\omega_2^2}{2}}\pm\Delta}\,,\label{wpm}
\end{eqnarray}
where
\begin{eqnarray}\Delta&=&\sqrt{\lambda_+^2(\omega_1+\omega_2)^2+
(\omega_1-\omega_2)^2[{\textstyle\frac{(\omega_1+\omega_2)^2}{4}}-\lambda_-^2]}\\
&=&\sqrt{{\textstyle\frac{(\omega_1^2-\omega_2^2)^2}{4}}+2\omega^2(\omega_1^2+\omega_2^2)}\,.
\end{eqnarray}
It is verified that for $\omega=0$ ($\lambda_{\pm}=0$),
$\Delta=\frac{\omega_1^2-\omega_2^2}{2}$ and hence
$\omega_{+(-)}=\omega_{1(2)}$, leading to $U_{jk}(t)=\delta_{jk}e^{-i\omega_j
t}$, $V_{jk}(t)=0$. The free evolution $a_j(t)=e^{-i\omega_j t}\,a_j$ is then
recovered. Also, in the special case  $\omega_1=\omega_2=\omega$ and $\lambda_+=\lambda_-=\kappa$
we recover the eigenfrequencies $\omega_{\pm}=\sqrt{\omega^2\pm 2\kappa \omega}$ of \cite{EKN.68}.

On the other hand, in the {\it isotropic} case $\omega_1=\omega_2=\omega_0$, we
have $\lambda_-=0$, implying  $\Delta=2\omega\omega_0$ and
\begin{equation}
\omega_{\pm}=\omega_0\pm\omega\,,\;\;\;\;V(t)=0\,,\;\;\;\;\;\;(\omega_1=\omega_2=\omega_0)
\,,\label{iso}
\end{equation}
which leads finally to
\begin{equation}
a_1(t)=e^{-i\omega_0 t}(a_1\cos\omega t +a_2\sin\omega t
)\,,\;\;a_2(t)=e^{-i\omega_0 t} (-a_1\sin\omega t + a_2\cos\omega t)\,.
 \label{isoa} \end{equation}
This is equivalent to a beam-splitter type transformation of angle $\omega t$
of the field operators. In this case  the angular momentum term commutes with
$H$ and just rotates the field operators with angular frequency $\omega$.

The general solution (\ref{8})--(\ref{solf}) can be derived by many other methods.
For instance, we may write the solution of Eq.\ (\ref{M}) for each operator as
\begin{equation}
a_j(t)=e^{iH t/\hbar}a_je^{-iHt/\hbar} =a_j+\frac{it}{\hbar}\left[H,a_j\right]
+\frac{1}{2!}\left(\frac{it}{\hbar}\right)^{2}\left[H,\left[H,a_j\right]\right]+\ldots
\label{9}
\end{equation}
which leads immediately to the form  (\ref{8}) for $a_j(t)$.  And insertion  of
a trial solution of the form (\ref{8}) in (\ref{7}) (i.e., the so called
Sen-Mandal approach \cite{SM.05,SM.13}) leads to a linear system of first order
differential equations for the coefficients $U_{jk}(t)$, $V_{jk}(t)$, namely
$i\,\dot{\cal U}={\cal H}{\cal U}$, i.e.,
\begin{equation}
i\left(\begin{array}{c}
\dot{\bm{U}}_{k}\\\dot{\bm{V}}_{k}^*\end{array}\right) = {\cal H}
\left(\begin{array}{c}
\bm{U}_{k}\\\bm{V}_{k}^*\end{array}\right)\,,\;\;\;k=1,2\label{8Me}
\end{equation}
where $\bm{U}_{k}=\left(\begin{array}{c}U_{1k}\\U_{2k}\end{array}\right)$,
$\bm{V}_{k}=\left(\begin{array}{c}V_{1k}\\V_{2k}\end{array}\right)$ are the
$k^{\rm th}$ column of $U$, $V$. Eq.\ (\ref{8Me}) and the initial conditions
$U_{jk}(0)=\delta_{jk}$, $V_{jk}(0)=0$ lead again to the solution
(\ref{sol0})--(\ref{solf}). We finally  notice that the system (\ref{7}) leads,
after successive derivation, to the fully decoupled quartic equations
\begin{equation}
\mathop{a_j}^{\cdot\cdot\cdot\cdot}
+(\omega_1^2+\omega_2^2+2(\lambda_+^2-\lambda_-^2))\ddot{a_j}
+((\lambda_++\lambda_-)^2-\omega_1\omega_2)
((\lambda_+-\lambda_-)^2-\omega_1\omega_2)a_j=0\label{o4}
\end{equation}
for $j=1,2$, which lead at once to the eigenfrequencies (\ref{wpm}) (after
inserting a trial solution $a_j(t)\propto e^{i\alpha t}$) and again to the
solution (\ref{sol0})--(\ref{solf}) after inserting the initial conditions for
the operators and their derivatives. The matrix ${\cal U}(t)$, and hence all
coefficients $U_{jk}(t)$, $V_{jk}(t)$, also satisfy Eq.\ (\ref{o4}).

\subsection{Dynamical stability and normal mode decomposition}
In the general case, a close inspection of the
eigenvalues (\ref{wpm}) reveals that $\omega_{\pm}$ are both {\it real} and
{\it non-zero} only if $\lambda_++\lambda_-<\sqrt{\omega_1\omega_2}$ or
$\lambda_+-\lambda_->\sqrt{\omega_1\omega_2}$,  i.e., if $\omega<\omega_2$ or
$\omega>\omega_1$,  which is equivalent to
\begin{equation}(\omega-\omega_1)(\omega-\omega_2)>0
 \label{st}\,.\end{equation}
Dynamical stability (bounded quasiperiodic dynamics) is then ensured if Eq.\
(\ref{st}) is satisfied. On the other hand, if
$\lambda_+-\lambda_-<\sqrt{\omega_1\omega_2}<\lambda_++\lambda_-$, i.e.,
$\omega_2<\omega<\omega_1$ or in general
\begin{equation}(\omega-\omega_1)(\omega-\omega_2)<0\label{inst}\,,\end{equation}
$\omega_+$ remains real but $\omega_-$ becomes {\it imaginary}
($\omega_-=i|\omega_-|$) implying that the dynamics becomes {\it unbounded}. In
this case we should just replace
\begin{equation}\frac{\sin\omega_- t}
{\omega_-}\rightarrow \frac{\sinh |\omega_-|\,t}{|\omega_-|}\,,\;\;
 \cos\omega_- t\rightarrow \cosh|\omega_-|\,t\end{equation}
in Eqs.\ (\ref{sol0})--(\ref{solf}), entailing that all operators ``increase''
(i.e., deviate from their initial values) exponentially with time.
Nevertheless, Eq.\ (\ref{Un}) and hence the commutation relations (\ref{com})
remain satisfied.

Finally, if $\omega=\omega_2$ or $\omega=\omega_1$, i.e., if
\begin{equation}(\omega-\omega_1)(\omega-\omega_2)=0\label{st0}\,,\end{equation}
we obtain a {\it critical} regime where  $\omega_+>0$ but $\omega_-=0$, in
which case the matrix ${\cal H}$ becomes  {\it non-diagonalizable} unless
$\omega_1=\omega_2$ (Landau case). If $\omega_-=0$, we should just replace the
corresponding  expressions by their natural limits,  i.e.,
\begin{equation}\frac{\sin\omega_- t}{\omega_-}\rightarrow t\,,
\;\;\cos\omega_- t\rightarrow 1 \,,
\label{rep3}\end{equation}
in Eqs.\ (\ref{sol0})--(\ref{solf}) (the ensuing solution also follows from
(\ref{exp}) after using the Jordan canonical form of ${\cal H}$ for
$\omega_-=0$ \cite{RK.09}), entailing that the dynamics is again unbounded if
$\omega_1\neq \omega_2$, with the deviation from the initial values increasing
now linearly with time.  Eqs.\ (\ref{Un}) are again preserved. In the Landau
case  $\omega=\omega_1=\omega_2$,  the coefficients of the linearly increasing
terms vanish and the dynamics is again bounded, given by Eq.\ (\ref{iso}) for
$\omega=\omega_0$.

A standard normal mode decomposition of $H$
becomes feasible in the dynamically stable phases ($\omega_{\pm}>0$)
\cite{RK.09}. In terms of the standard normal mode boson  operators $b_{\pm}$,
$b^\dagger_{\pm}$ given in the appendix, we may express (\ref{1}) in the first
dynamically stable sector $\omega<\omega_2$ as
\begin{equation}
H=\hbar\omega_+
(b^\dagger_+b_++{\textstyle\frac{1}{2}})+\hbar\omega_-(b^\dagger_-b_-+{\textstyle\frac{1}{2}})\,,
\;\;\omega<\omega_2\label{r1}\end{equation}
with $b_{\pm}(t)=e^{-i\omega_\pm t}b_\pm(0)$, whereas in the second dynamically
stable sector $\omega>\omega_1$ we have
\begin{equation}H=\hbar\omega_+ (b^\dagger_+b_++{\textstyle\frac{1}{2}})
-\hbar\omega_-(b^\dagger_-b_-+{\textstyle\frac{1}{2}})\,,\;\;\omega>\omega_1\,,
 \label{r2}\end{equation}
with $b_{-}(t)=e^{i\omega_- t}b_-(0)$ (and $\omega_->0$).  Eq.\ (\ref{r2})
entails that in this  region, the system is no longer energetically stable.

\section{Application}
\subsection{Squeezing and entanglement}
We have now all the elements for
investigating the evolution of distinct quantum properties of the system, such
as entanglement and squeezing. We start by noting that the number operators for
each mode are given by (here $j,k,l=1,2$)
\begin{eqnarray}N_j(t)\equiv a^\dagger_j(t)a_j(t)&=&\sum_{k,l}
[U_{jk}^*(t)U_{jl}(t)a^\dagger_ka_l+V_{jk}^*(t)V_{jl}(t)a_ka^\dagger_l\nonumber\\&&
+U_{jk}^*(t)V_{jl}(t)a^\dagger_ka^\dagger_l+V_{jk}^*(t)U_{jl}(t)a_k a_l]\,,
 \label{34}\end{eqnarray}
indicating that they will acquire a non-zero average even if there are
initially no bosons: If the system starts at the separable vacuum
$|00\rangle\equiv |0_1\rangle|0_2\rangle$, where $a_j|0_j\rangle=0$,  from
Eqs.\ (\ref{34}) and (\ref{sol0})--(\ref{solf}) we obtain, setting $\langle
O\rangle_0\equiv \langle 00|O|00\rangle$ and $j,k=1,2$,
\begin{eqnarray}\langle N_j(t)\rangle_0&=&
 \sum_{k}|V_{jk}(t)|^2\label{V2}\\
&=&\frac{\omega^2(\omega_1-\omega_2)^2}{16\omega_1\omega_2}
\left[\left|\sum_{\nu=\pm}{\textstyle(1+\nu\frac{(\omega_1-\omega_2)^2}{2\Delta})\frac{\sin\omega_\nu
t}{\omega_\nu}+i\nu\frac{(\omega_1-\omega_2)\cos\omega_\nu
t}{\Delta}}\right|^2\right.
\nonumber\\
&&+\left.\frac{\omega^2(\omega_1+\omega_2)^2\omega_1\omega_2}{\omega_j^2|\Delta|^2}\left|{\textstyle
\frac{\sin\omega_+ t}{\omega_+}-\frac{\sin\omega_-
t}{\omega_-}}\right|^2\right] \,,\label{V2a}
\end{eqnarray}
which will be normally non-zero for $t>0$ unless $\omega_1=\omega_2$.  $\langle
N_j(t)\rangle$ is then proportional to $(\omega_1-\omega_2)^2$, and is larger
in the mode with the lowest frequency $\omega_j$ (due to the last term in
(\ref{V2a})).

If the initial state is instead a product
$|\alpha_1\alpha_2\rangle\equiv|\alpha_1\rangle|\alpha_2\rangle$ of {\it
coherent} states $|\alpha_{j}\rangle$ for each oscillator, with
$a_j|\alpha_j\rangle=\alpha_j|\alpha_j\rangle$, the same expressions (\ref{V2})--(\ref{V2a})
remain valid for the corresponding {\it covariance} of the operators
$a_j^\dagger(t)$, $a_j(t)$:
\begin{eqnarray}\langle N_j(t)\rangle_\alpha
-\langle a^\dagger_j(t)\rangle_\alpha\langle a_j(t)\rangle_\alpha&=&\langle N_j(t)\rangle_0=
 \sum_{k}|V_{jk}(t)|^2\,,\label{V3}\end{eqnarray}
where $\langle O\rangle_\alpha\equiv \langle
\alpha_1\alpha_2|O|\alpha_1\alpha_2\rangle$. Eq.\ (\ref{V3}) is then {\it
independent} of $\alpha_1$ and $\alpha_2$.

Similarly, we may evaluate the coordinates and momenta fluctuations and their
dimensionless ratios to their initial values,
\begin{eqnarray}
R^2_{Q_j}(t)&=&\frac{\langle Q_j^2(t)\rangle_\alpha-
\langle {Q}_j(t)\rangle_\alpha^2}{\langle Q_j^2(0)\rangle_\alpha
-\langle {Q}_j(0)\rangle_\alpha^2}\,,\;\;
R^2_{P_j}(t)=\frac{\langle P_j^2(t)\rangle_\alpha-\langle {P}_j(t)\rangle_\alpha^2}
{\langle P_j^2(0)\rangle_\alpha-\langle {P}_j(0)\rangle_\alpha^2}\,,\label{RQP}
\end{eqnarray}
which for a coherent initial state satisfy $R_{Q_j}(t)R_{P_j}(t)\geq 1$,  due
to the uncertainty principle and the fact that a coherent initial state has
minimum uncertainty. Squeezing in $Q_j$ or $P_j$ occurs whenever $R_{Q_j}(t)$
or $R_{P_j}(t)$) becomes smaller than $1$. We obtain, explicitly,
\begin{eqnarray}
R^2_{Q_j(P_j)}(t)&=&1+2[\langle N_j(t)\rangle_\alpha-|\langle a_j(t)\rangle_\alpha|^2
\pm{\rm Re}(\langle a_j^2(t)\rangle_\alpha-\langle a_j(t)\rangle_\alpha^2)]\nonumber\\
&=&1+2[\langle N_j(t)\rangle_0\pm{\rm Re}(\langle a_j^2(t)\rangle_0)]\,,
 \label{RQPs}\end{eqnarray}
where ${\rm Re}$ denotes real part,  $+$ ($-$) corresponds to $Q_j$
($P_j$) and
\begin{equation}
\langle a_j^2(t)\rangle_0=\sum_{k}U_{jk}(t)V_{jk}(t)\,.
\label{aj2}
\end{equation}
These ratios are then also independent of $\alpha_1,\alpha_2$, and
deviate from $1$ unless $V(t)=0$.

An initial coherent state is a pure separable gaussian state,  which under the
present Hamiltonian will remain gaussian (but no longer separable) $\forall t$.
Its entanglement entropy $S(t)$ can then be evaluated through the gaussian
state formalism \cite{RR.11,RCR.14,AE.02,AS.04,BL.05}  and can be written as
\begin{equation}S(t)=-{\rm Tr}\, \rho_j(t)\ln\rho_j(t)
=-f(t)\ln f(t)+[1+f(t)]\ln [1+f(t)]\,,\label{S}
\end{equation}
where $\rho_j(t)$ denotes the reduced state of one of the modes and $f(t)$ is
the symplectic eigenvalue of the single mode covariance matrix:
\begin{eqnarray}
f(t)&=&\sqrt{(\langle N_j(t)\rangle_\alpha-|\langle a_j(t)\rangle_\alpha|^2+{\textstyle\frac{1}{2}})^2
-|\langle a_j^2(t)\rangle_\alpha-\langle a_j(t)\rangle_\alpha^2|^2}-{\textstyle\frac{1}{2}}\nonumber\\
&=&\sqrt{(\langle N_j(t)\rangle_0+{\textstyle\frac{1}{2}})^2
-|\langle a_j^2(t)\rangle_0|^2}-{\textstyle\frac{1}{2}}\label{ft}
\end{eqnarray}
which is non-negative and the same for $j=1$ or $j=2$ if the state is pure and
gaussian. It represents the effective occupation number of the mode
\cite{RCR.14}. It is obviously also independent of $\alpha_1,\alpha_2$, i.e.,
the same for {\it any} coherent initial state. The entanglement entropy
(\ref{S}) is just an increasing concave function of $f(t)$. Again, in the isotropic
case $\omega_1=\omega_2$,  $V(t)=0$ (Eq.\ (\ref{iso})), entailing no generated
entanglement when starting from $|\alpha_1\alpha_2\rangle$.

\subsection{Results}

\begin{figure}[ht!]
\vspace*{0cm}

\centerline{\hspace*{-0.2cm}\scalebox{.8}{\includegraphics{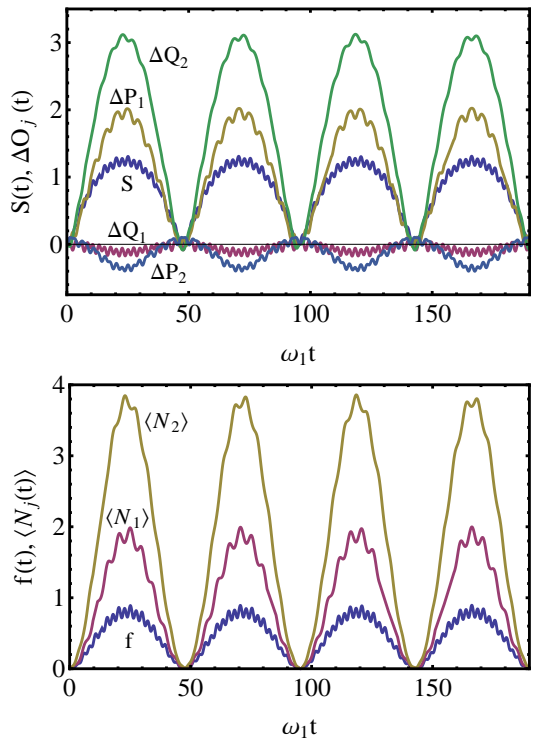}}}
\caption{(Color online) Top: Temporal evolution of the entanglement entropy
$S(t)$ (\ref{S}) and the shifted squeezing ratios (\ref{Delta}) for the
operators $Q_j$, $P_j$ of each oscillator, starting from a separable coherent
state. We have set here $\omega_2=\omega_1/2$  with a rotation frequency
$\omega=0.49\omega_1$, such that the system is close to the first instability
(occurring when $\omega$ reaches $\omega_2$). Both $Q_1$ and $P_2$ exhibit
appreciable squeezing ($\Delta O_j(t)<0$), whose evolution is in phase with
that of entanglement. Bottom: The corresponding symplectic eigenvalue $f(t)$
(\ref{ft}) determining the entanglement entropy, and the average boson numbers
of each mode when starting from the separable vacuum (or the covariances
(\ref{V3}) when starting from a coherent state). Quantities plotted are
dimensionless. } \label{f1}
\end{figure}

\begin{figure}[ht!]
\vspace*{0cm}

\centerline{\hspace*{-0.2cm}\scalebox{.8}{\includegraphics{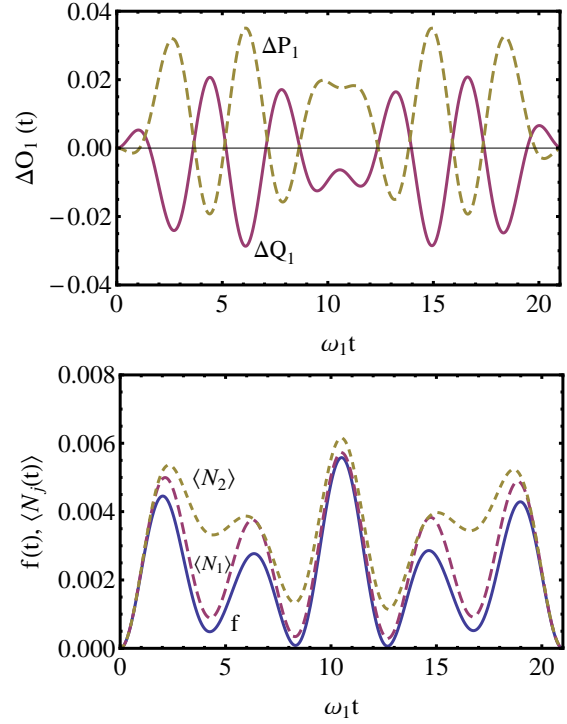}}}
\caption{(Color online) Top: The shifted squeezing ratios for $Q_1$ and $P_1$
for weak coupling $\omega=0.15 \omega_1$. Now both $Q_j$ and $P_j$ exhibit
small alternating squeezing. Bottom: The corresponding value of $f(t)$
(\ref{ft}), together with the average boson number of each oscillator. The behavior of
the entanglement entropy $S(t)$ (\ref{S}) is similar to that of $f(t)$. }
\label{f2}
\end{figure}

Results for the previous quantities are depicted in Figs.\ \ref{f1}--\ref{f2}
for $\omega_2<\omega_1$. We concentrate on the first dynamically stable sector
$\omega<\omega_2$. For improved visualization of squeezing, we use there the
quantities
\begin{equation}\Delta Q_j(t)\equiv R_{Q_j}(t)-1\,,\;\;\Delta P_j(t)\equiv R_{P_j}(t)-1\,,
\label{Delta}\end{equation}
with squeezing in $Q_j$ $(P_j)$ indicated by a {\it negative} value of $\Delta
Q_j$ ($\Delta P_j$).

In Fig.\ \ref{f1} we consider the anisotropic case  $\omega_2=\omega_1 /2$
with  $\omega$ close to $\omega_2$, i.e., to  the first instability, such that
$\omega_+\approx 1.31\omega_1$, $\omega_-\approx 0.07\omega_1$. The evolution
then exhibits large amplitude low frequency oscillations governed by
$\omega_-$,  together with small amplitude high frequency oscillations governed
by $\omega_+$. The picture clearly shows that in this regime squeezing and
entanglement oscillate {\it in phase}: Maximum entanglement occurs at times
$t_n\approx n\pi/(2\omega_-)$, $n$ odd (see Eqs.\ (\ref{ap1})--(\ref{ap2})),
simultaneously with maximum squeezing in the operators $Q_1$ and $P_2$, and
maximum average boson number in the oscillators (or maximum covariance
(\ref{V3}) in the case of an initially coherent state).

These results can be approximately described  by conserving just the main terms
in $V_{jk}(t)$ and $U_{jk}(t)$ (Eqs.\ (\ref{sol0})--(\ref{solf})) for small
$\omega_-$, which are those proportional to $\omega_-^{-1}$.  We obtain
\begin{eqnarray}
\langle N_j(t)\rangle_0&\approx&{\textstyle\frac{\omega^2(\omega_1-\omega_2)^2}
{16\omega_1\omega_2}[(1-\frac{(\omega_1-\omega_2)^2}{2\Delta})^2+
\frac{\omega^2(\omega_1+\omega_2)^2\omega_1\omega_2}{\omega_j^2\Delta^2}]
\frac{\sin^2\omega_- t}{\omega_-^2}} \,,\label{ap1}\\
\langle a_j^2(t)\rangle_0&\approx&(-1)^j
{\textstyle\frac{\omega^2(\omega_1^2-\omega_2^2)} {16\omega_1\omega_2}
[\frac{2\omega_1\omega_2(1-\delta_j)}{\Delta}-(1-\frac{(\omega_1-\omega_2)^2}{2\Delta})
(1-\frac{(\omega_1+\omega_2)^2}{2\Delta})]\frac{\sin^2\omega_- t}{\omega_-^2}}
\,, \nonumber\\&&\label{ap2}
\end{eqnarray}
where  $\delta_j$ is given in (\ref{delj}). Hence, both $\langle
N_j(t)\rangle_0$ and $\langle a_j^2(t)\rangle_0$ become proportional to
$\sin^2\omega_- t$, entailing that all quantities plotted in Fig.\ \ref{f1}
will be governed by this term, thus oscillating in phase. The presence of a
factor $\omega_j^{-2}$ in (\ref{ap1}) also entails $\langle N_2(t)\rangle_0\geq
\langle N_1(t)\rangle_0$, i.e., the boson number will be larger in the mode
with the smallest frequency, as verified in the bottom panel of Fig.\ \ref{f1}.

Moreover, for $\omega$ below but close to $\omega_2$, the bracket in
(\ref{ap2}) will be positive for $j=1,2$ and larger for $j=2$ (since
$0<\delta_2<\delta_1<1$), leading  to $\langle a_1^2(t)\rangle_0\leq 0$ and
$\langle a_2^2(t)\rangle_0\geq 0$, with $|\langle a_1^2(t)\rangle_0|\leq
|\langle a_2^2(t)\rangle_0|$. Hence, according to Eq.\ (\ref{RQPs}), squeezing
will occur in $Q_1$ and $P_2$, as seen in Fig.\ \ref{f1}, being more pronounced
in $P_2$. Eqs.\ (\ref{ap1})--(\ref{ap2}) also show that squeezing and
entanglement are driven by the anisotropy, i.e.,  they vanish for
$\omega_1=\omega_2$ and at fixed $\omega$ become larger as $\omega_1-\omega_2$
increases (or the ratio $\omega_2/\omega_1$ decreases). They also increase, of
course, as $\omega$ approaches $\omega_2$,  i.e., as the first instability is
reached ($\omega_-=0$ for $\omega=\omega_2$).  We  also mention that for
$\omega$ above but close to $\omega_1$ (i.e., in the second stable region but
close to instability), the behavior is similar although squeezing will occur
for $P_1$ and $Q_2$, since the bracket in (\ref{ap2}) becomes negative (with
$\delta_2>\delta_1>1$), implying  $\langle a_1^2(t)\rangle_0\geq 0$, $\langle
a_2^2(t)\rangle_0\leq 0$.

In contrast, away from instability ($\omega$ well below $\omega_2$), quantum
effects become much smaller even though they remain non-zero, as seen in Fig.\
\ref{f2} for $\omega=0.15\omega_1$ (where $\omega_+\approx 1.04\omega_1$,
$\omega_-\approx 0.45\omega_1$). Alternating squeezing in both $P_j$ and $Q_j$
is now observed (the behavior of $\Delta P_2$ and $\Delta Q_2$ is analogous)
and the correspondence with the evolution of entanglement (i.e. with $f(t)$) is
less direct, with the maxima of $f(t)$  reflecting essentially the largest
squeezing (that of $Q_1$ or $P_2$). Nonetheless, the average boson numbers
$\langle N_j(t)\rangle$ follow approximately $f(t)$.

We can easily understand these results by considering the expansion of the
exact expressions (\ref{V2})--(\ref{aj2}) for small $\omega$ ($|\omega|\ll {\rm
Min}[\omega_1,\omega_2]$). We obtain, neglecting terms of order $\omega^4$,
\begin{eqnarray}\langle N_j(t)\rangle_0&\approx&
\frac{\omega^2(\omega_1-\omega_2)^2}{\omega_1\omega_2(\omega_1 + \omega_2)^2}
\sin^2{\textstyle\frac{\omega_1 + \omega_2}{2}} t\,,\label{Nj2a}\\
\langle a_j^2(t)\rangle_0&\approx&i(-1)^{j+1}\frac{\omega^2(\omega_1^2-\omega_2^2)}
{2\omega_1\omega_2\omega_j}e^{-i\frac{\omega_1+\omega_2}{2}t}[e^{-i\omega_j t}
{\textstyle\frac{\sin\frac{\omega_1-\omega_2}{2}t}{\omega_1-\omega_2}-
\frac{\sin\frac{\omega_1+\omega_2}{2}t}{\omega_1+\omega_2}}]\label{aj2a2}\,,
\end{eqnarray}
which show that {\it both} $\langle N_j(t)\rangle_0$ and $\langle
a_j^2(t)\rangle_0$ are of order $\omega^2$, with $\langle N_1(t)\rangle_0=\langle
N_2(t)\rangle_0$ at this order. This implies that $|\langle
a_j^2(t)\rangle_0|^2$ will be of order $\omega^4$, so that Eq.\ (\ref{ft})
leads to $f(t)\approx \langle N_j(t)\rangle_0$ up to $O(\omega^2)$. Hence, the
$\langle N_j(t)\rangle_0$ and $f(t)$ will be close for small $\omega$.

Besides, ${\rm Re}[\langle a_j^2(t)\rangle_0]$ will change its sign as $t$
evolves, indicating that squeezing will alternate between $Q_j$ and $P_j$,
being again larger for the oscillator with the lowest frequency due to the
factor $\omega_j^{-1}$ in (\ref{aj2a2}). Note that for small $\langle
N_j(t)\rangle_0$ and $\langle a_j(t)\rangle_0$, $R_{Q_j(P_j)}(t)\approx
1+\langle N_j(t)\rangle_0\pm {\rm Re}[\langle a_j(t)^2\rangle_0 ]$, which will
not strictly follow $\langle N_j(t)\rangle_0$, as $\langle a_j^2(t)\rangle_0$
is of the same order as $\langle N_j^2(t)\rangle_0$ but not proportional to it.
Finally, it is verified from (\ref{Nj2a})--(\ref{aj2a2}) that $\langle
N_j(t)\rangle_0$ and $\langle a_j^2(t)\rangle_0$ are again proportional to
$(\omega_1-\omega_2)^2$ and $\omega_1^2-\omega_2^2$ respectively,  hence
vanishing for $\omega_1=\omega_2$ and leading to a larger entanglement and
squeezing as the anisotropy $\omega_1-\omega_2$ increases.

We finally mention that for short times $t$ such that $\omega_{\pm} t\ll 1$,
we obtain, after an expansion up to $O(t^2)$ of the exact expressions
(\ref{V2})--(\ref{aj2}),
\begin{eqnarray}\langle N_j(t)\rangle_0&\approx&
\frac{\omega^2(\omega_1-\omega_2)^2\,t^2}{4\omega_1\omega_2}\,,\label{Nj3a}\;\;\;
\langle a_j^2(t)\rangle_0\approx(-1)^{j+1}\frac{\omega^2(\omega_1^2-\omega_2^2)
\,t^2}{4\omega_1\omega_2}\,,
\end{eqnarray}
which are in agreement with the $t\rightarrow 0$ limits of Eqs.\
(\ref{Nj2a})--(\ref{aj2a2}). Hence, these quantities increase initially {\it
quadratically} with time $t$, with $\langle N_1(t)\rangle_0=\langle
N_2(t)\rangle_0 \approx f(t)$ in this limit. Moreover, in this regime $\langle
a_1^2(t)\rangle_0=-\langle a_2^2(t)\rangle_0$ is real, and positive if
$\omega_1>\omega_2$. This  entails that squeezing will initially  start in
$P_1$ (as seen in Fig.\ \ref{f2}) and $Q_2$, with
\begin{equation}
R_{P_1}(t)\approx R_{Q_2}(t)\approx 1-\frac{\omega^2(\omega_1-\omega_2)\omega_2}{2\omega_1\omega_2}t^2\,.
\end{equation}

\section{Conclusions}
We have derived the exact analytical closed form solution for the field
operators of two linear oscillators coupled through angular momentum. We then
applied the solution to investigate the relation between squeezing and
entanglement generation in this model when starting from a separable coherent
state. In the vicinity of  instability,  the generated entanglement between the
modes shows a large amplitude--low frequency behavior (almost periodic), which
is reflected in a similar behavior of the squeezing in the coordinate of one of
modes  and the momentum of the other mode. A different behavior occurs in the
weak coupling regime, away from instability, where the generated entanglement
is small and the  squeezing is weak, exhibiting an  essentially alternating
behavior for  the coordinate and momentum of each oscillator. Approximate
analytical expressions  describing these two regimes have also been derived
from the general exact solution.

The present solution has of course potential applications for studies of other
quantum statistical properties such as  higher order squeezing, antibunching of
photons and other nonclassical photon statistics. The solution  is also of
interest for quantum information applications.  As stated in the introduction,
the present model admits  distinct physical realizations, so that results could
in principle be tested in quite different scenarios (optical simulations,
particles in anisotropic harmonic traps, condensates, etc.). We remark, finally, that
expressions similar to (\ref{M})--(\ref{exp}) and (\ref{V2}), (\ref{aj2})
remain formally valid for general systems of $n$ harmonic modes interacting
through quadratic (in $a_j$, $a^\dagger_j$) couplings, replacing ${\cal H}$ by
the corresponding $2n\times 2n$ matrix.

\textbf{Acknowledgements:} \emph{We are thankful to the Third World Academy of
Sciences (TWAS), Trieste, Italy and CONICET of Argentina, for financial support
through TWAS-UNESCO fellowship program. NC and RR acknowledge support from
CONICET and CIC of Argentina, while SM thanks the University Grants Commission,
Government of India, for support through the research project
(F.No.42-852/2013(SR)). }

\section*{Appendix}
By means of the canonical transformation ($j=1,2$)
\begin{equation}P_j'=P_j+\gamma Q_{3-j},\;Q'_j=(Q_j-\eta P_{3-j})/(1+\eta\gamma)
\end{equation}
where $\gamma=\frac{2\Delta-\omega_1^2+\omega_2^2}{4\omega}$,
$\eta=\frac{2\gamma}{\omega_1^2+\omega_2^2}$, we may rewrite (\ref{1})  in the
decoupled form (we set here $m=1$)
\begin{eqnarray}
H&=&\frac{1}{2}\sum_{j=1,2}(\alpha_j{P'_j}^2+\beta_j {Q'_j}^2)\,,\label{nm1}
\end{eqnarray}
where $\alpha_j=1-\frac{\omega}{\Delta}(\gamma+(-1)^j\omega)$,
$\beta_j=\frac{\Delta}{\omega}(\gamma-(-1)^j\omega)$. Here $\alpha_j>0$,
$\beta_j>0$ for $j=1,2$ in the fully stable region $\omega<\omega_2$, whereas
$\alpha_{1}>0$, $\beta_{1}>0$, $\alpha_2<0$, $\beta_2<0$  in the second
dynamically stable sector $\omega>\omega_1$, with $\alpha_2>0$, $\beta_2<0$ in
the unstable sector $\omega_2<\omega<\omega_1$ and $\beta_2=0$ at the borders
$\omega=\omega_2$ or $\omega=\omega_1$. Eq.\ (\ref{nm1}) then leads, in the
dynamically stable regions $\omega<\omega_2$ or $\omega>\omega_1$, to
\begin{eqnarray}
H&=&\hbar\omega_+\frac{P_+^2+Q_+^2}{2}\pm\hbar\omega_-\frac{P_-^2+Q_-^2}{2}\,,\label{Hd}
\end{eqnarray}
where $\omega_{\pm}=\sqrt{\alpha_{^1_2}\beta_{^1_2}}$ (real),
$P_{\pm}=\sqrt[4]{\alpha_{^1_2}/\hbar\beta_{^1_2}}\,P_{^1_2}$, $Q_{\pm}
=\sqrt[4]{\beta_{^1_2}/\hbar\alpha_{^1_2}}\,Q_{^1_2}$, and the minus sign in
(\ref{Hd}) applies  for $\omega>\omega_1$. Eqs.\ (\ref{r1})--(\ref{r2}) are
then obviously obtained for $b_{\pm}=(Q_{\pm}+iP_{\pm})/\sqrt{2}$,
$b^\dagger_{\pm}=(Q_{\pm}-iP_{\pm})/\sqrt{2}$. The possible
normal representations in the unstable regime are discussed in detail in
\cite{RK.09,RK.05}.


\section*{References}

\begin{thebibliography}{999}

\bibitem{Louisell} Louisell W H 1990 \emph{Quantum Statistical
Properties of Radiation}, Wiley (NY)
\bibitem{Walls}Walls D F and Milburn G J 1994 \emph{Quantum Optics},
Springer (Berlin)
\bibitem{Mollow} Mollow B R \PR 1967 \textbf{162} 1256
\bibitem{EKN.68} Estes L E, Keil T H and Narducci L M 1968 \PR {\bf 175} 286
\bibitem{Iafrate}Iafrate G J and Croft M 1975 \PR A  \textbf{12}  1525
\bibitem{Holzwarth} Holzwarth G and  Chabay I 1972  \emph{J.\ Chem.\ Phys.\ }
\textbf{57} 1632
\bibitem{Belkin} Belkin M A, Shen Y R and Flytzanis C 2002
\emph{Chem.\ Phys.\ Lett.} \textbf{363} 479
\bibitem{Cochrane} Cochrane P T, Milburn G J and Munro W J 2000 \PR A
 \textbf{62}  062307
\bibitem{Fan} Fan H,  Li C and  Jiang Z 2004 \PL A
 \textbf{327}  416
\bibitem{Ng} Ng K M,  Lo C F 1997 \PL A  \textbf{230} 144
\bibitem{Va.56}  Valatin J G 1956 \emph{Proc.\ R.\ Soc.} London {\bf 238} 132
\bibitem{FK.70}  Feldman A and Kahn A H 1970 \PR B {\bf 1}  4584
\bibitem{RS.80} Ring P and Schuck P 1980 {\emph The Nuclear Many-Body Problem,}
Springer (NY); \\Blaizot J P  Ripka G 1986  {\emph Quantum Theory of Finite
Systems,} MIT Press (MA)
\bibitem{MC.94}Madhav A V and  Chakraborty T 1994 \PR B {\bf 49} (1994) 8163
\bibitem{LNF.01} Linn M, Niemeyer M and Fetter A L 2001
 \PR A {\bf 64}  023602
\bibitem{OO.04}   Oktel M \"O 2004 \PR A {\bf 69} 023618
\bibitem{AF.07}  Fetter A L  2007 \PR A {\bf 75}  013620
\bibitem{ABL.09}  Aftalion A,  Blanc X and Lerner N 2009 \PR A
 {\bf 79} 011603(R)
\bibitem{ABD.05} Aftalion A,  Blanc X and  Dalibard J 2005 \PR A {\bf 71}  023611
\bibitem{BDZ.08}  Bloch I,  Dalibard J and  Zwerger W 2008 \RMP {\bf 80}  885
\bibitem{FF.09} Fetter A L  2009 \RMP {\bf 81} 647
 \bibitem{PE.94} P\v{e}rina J, Hradil Z and Jur\v{c}o B 1994
{\it Quantum Optics and Fundamentals of Physics}, Kluwer, Dordrecht
\bibitem{RK.09} Rossignoli R and Kowalski A M 2009 \PR A {\bf 79} 062103
\bibitem{RR.11} Reb\'on L and Rossignoli R  2011 \PR A  {\bf 84} 052320
\bibitem{RCR.14} Reb\'on L, Canosa N and Rossignoli R 2014 \PR A {\bf 89} 042312
\bibitem{SLRD.13} Schachenmayer J et al
2013 \PR X {\bf 3}  031015;
Daley A J et al 2012
\PRL {\bf 109} 020505; Bardarson J H, Pollmann F and  Moore J E 2012
\PRL {\bf 109} 017202
\bibitem{NC.00} Nielsen M and  Chuang I L 2000 {\it Quantum Computation and
        Quantum Information} (Cambridge Univ.\ Press, Cambridge, UK)
\bibitem{DS.01}  S\o rensen A et al 2001 {\it Nature} {\bf 409} 63;
Orzel C et al 2001 \emph{Science} {\bf 291}  2386; Bigelow N 2001 \emph{Nature}
 {\bf 409}  27
\bibitem{JE.08} Esteve J et al 2008 \emph{Nature} {\bf 455}  1216
\bibitem{GT.09} G\"uhne O and Toth G 2009 {\it Phys.\ Rep.\ } {\bf 474} 1
\bibitem{Ma.11} Ma J, Wang X, Sun C P, Nori F 2011  {\it Phys. Rep.} {\bf 509} 89
\bibitem{CB.05} Choi S and Bigelow N P 2005 \PR A {\bf 72} 033612
\bibitem{TO.09} Toth G, Knapp C, G\"uhne O and Briegel H J, 2009 \PR A {\bf 79}  042334.
\bibitem{CH.10}  Chung N N et al 2010 \PR A {\bf 82} 014101;
Chew L C and  Chung N N 2014 \emph{Symmetry} {\bf 6}  295
\bibitem{CG.12} Gross C 2012 \JPB {\bf 45}  103001
\bibitem{SM.05} Sen B and Mandal S 2005 \emph{J.\ Mod.\ Opt} {\bf 52}  1789;
 Sen B, Mandal S and P\v{e}rina J 2007 \JPB {\bf 40}  1417
\bibitem{SM.13}  Sen B et al 2013 \PR A {\bf 87} 022325
 \bibitem{AE.02}  Audenaert K et al 2006
\PR A {\bf 66} 042327
 \bibitem{AS.04} Adesso G, Serafini A and Illuminati F 2004
\PR A {\bf 70}, 022318; Serafini A, Adesso G and
Illuminati F 2005 \PR A {\bf 71} 032349
\bibitem{BL.05} Braunstein S L and  van Loock P 2005
 \RMP {\bf 77}  513; Weedbrook  C et al 2012 \RMP {\bf 84} 621
\bibitem{RK.05} Rossignoli R and Kowalski A M 2005 \PR A {\bf 72} 032101

\end{thebibliography}
\end{document}